# Structural Superlubricity at High Sliding Speeds under Ambient Conditions

Wai H. Oo[1], Paul D. Ashby[2*], Mehmet Z. Baykara[1*]

[1]Department of Mechanical Engineering, University of California Merced, Merced, CA 95343, USA

[2]Molecular Foundry, Lawrence Berkeley National Laboratory, Berkeley, CA 94720, USA

*Corresponding authors: pdashby@lbl.gov, mehmet.baykara@ucmerced.edu

Structural superlubricity is an intriguing physical phenomenon, whereby sliding at a structurally incommensurate, atomically flat interface yields vanishingly small friction forces. Despite its recent experimental validation, critical questions remain regarding the physical limitations of the concept. In particular, it is not known whether the ultra-low friction state would persist at high sliding speeds relevant for practical, small-scale mechanical systems. Here, we perform sliding experiments via atomic force microscopy on gold nanoislands on graphite at increasing speeds, extracting interfacial friction forces under ambient conditions. A heterodyne detection methodology enables the extraction of extremely weak friction signals buried deep in the noise, revealing that the structurally superlubric regime extends over 100 μm/s with minimal changes in friction force, spanning three orders of magnitude in sliding speed. Our results contribute significantly to the pursuit of functional, superlubric mechanical devices.

Structural superlubricity is a fascinating physical state where friction between two surfaces in relative motion nearly vanishes, a result of structural incommensurability at the interface. The idea was first hypothesized in 1990, when Hirano and Shinjo (*1*) theoretically predicted that the friction force should practically disappear at an interface formed by two atomically flat, molecularly clean, weakly interacting, and structurally incommensurate surfaces, because the potential energy barrier to sliding that each slider atom experiences vanishes with increasing the size of the slider, Fig. 1. This consequently results in a sub-linear dependency of friction force on contact area, leading to ultra-low frictional resistance to sliding (*2*).

This prediction of an ultra-low friction state that arises intrinsically, i.e., without the use of any lubricants, has implications of mechanical systems with minimal energy dissipation that is both intriguing and promising from a practical point of view. As such, following the publication of the theoretical ideas, research has been focused on observing structural superlubricity in laboratory experiments. In 2004, using a variation of atomic force microscopy (AFM), Dienwiebel *et al.* provided the first clear, quantitative experimental evidence of structural superlubricity by sliding a graphene flake on graphite in structurally commensurate and incommensurate configurations (*3*). In 2013, Dietzel *et al.* verified the theoretically predicted scaling laws of structural superlubricity by way of AFM-based sliding experiments performed on antimony and gold nanoislands on graphite, under ultrahigh vacuum (UHV) conditions (*4*). A few years later, Cihan *et al.* revealed that the observation of structural superlubricity is not confined to the pristine, yet impractical, UHV environment, by way of ultra-low friction forces recorded during the sliding of gold nanoislands on graphite under ambient conditions, in alignment with theoretically predicted scaling laws (*5*). These results were later extended to nanoislands made of another noble metal, platinum (*6*), which were also utilized to show that partial interfacial oxidation leads to a weakening but not a total breakdown of the structurally superlubric state (*7*). In parallel with this body of work focusing on sub-micrometer-scale contacts between metal sliders and graphite substrates, significant progress was achieved in the extension of structural superlubricity to the micrometer length scale, e.g. at interfaces formed between two-dimensional materials (*8*) and graphite counter-surfaces (*9*). Overall, the milestones summarized here clearly demonstrate that structural superlubricity continues to advance from theoretical conjecture to a concept that has real potential for practical applications in mechanical systems.

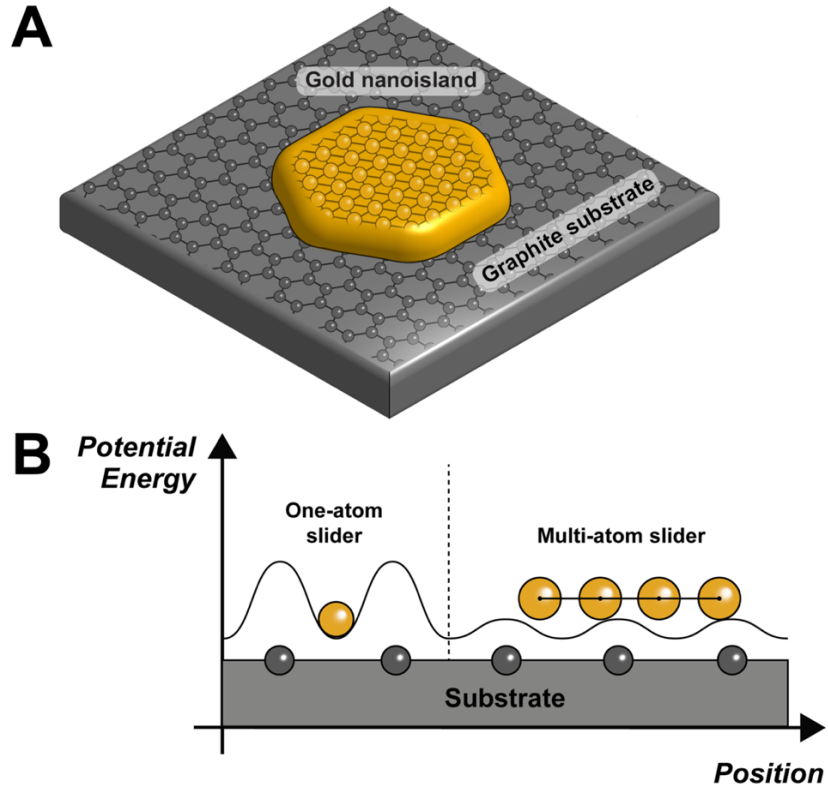

**Fig 1. Basic concept of structural superlubricity.** (**A**) Cartoon model of our structurally superlubric system, gold nanoislands on graphite. The structural incommensurability between the gold (111) nanoisland surface and the graphite (0001) substrate surface arises due to lattice mismatch. (**B**) Schematic illustration of the effect of structural incommensurability on the potential energy barrier per sliding atom: as the number of slider atoms increases, the energy barriers to sliding decrease. For large sliders, the effective barrier height approaches zero, leading to vanishing friction forces despite increasing contact size.

In order to realize the potential of structural superlubricity for mechanical applications, many questions about the robustness of the phenomenon against varying physical factors and operating conditions need to be answered. Perhaps the most important of these questions is how the miniscule friction forces that are experienced in structural superlubricity would evolve with increasing sliding speed. The criticality of this question for engineering applications arises due to the fact that sliding speeds in small-scale mechanical systems (~100 µm/s and above) are orders of magnitude higher than those in typical AFM experiments employed to study structural superlubricity (which are typically on the order of 1 µm/s and below) (*10*). In particular, if friction forces indeed increase strongly with sliding speed in a Stokesian manner as predicted by some theoretical work (*11–13*), this could essentially result in a breakdown of the superlubric state when speeds relevant for practical mechanical systems are reached, effectively making the pursuit of functional, structurally superlubric mechanical systems futile. While small friction forces claimed to arise due to structural superlubricity have been reported for the high-speed "self-retraction" phenomenon observed in graphite counter-surfaces as inferred from optical measurements (*14*), no experimental studies of friction forces as a direct function of sliding speed have been performed yet.

Motivated as above, we present here AFM-based sliding experiments on gold nanoislands situated on graphite, to investigate the evolution of friction forces with increasing sliding speed in structural superlubricity. A series of speed-dependent AFM experiments, spanning three orders of magnitude in sliding speed (from 0.6 µm/s to 115 µm/s), performed on a total of 10 gold nanoislands, show a negligible increase in friction forces with sliding speed, demonstrating the remarkable robustness of the structurally superlubric state against drastic changes in this critical parameter.

The material system studied here consists of gold nanoislands of various size and shape on graphite (for synthesis and characterization details, please see Materials and Methods in Supplementary Materials). The AFM-based sliding experiments performed on this material system were conducted in the "tip-on-top" mode (Fig. 2) (*15*). While details of the experimental methodology may be found in Materials and Methods, the basic procedure initially involves approaching and establishing contact with the top surface of a suitable nanoisland (located on a graphite terrace, away from defects and step edges) with the AFM tip. Once contact is established (under the influence of adhesive forces only, which were typically on the order of 10 nN and below), the sample is laterally modulated by way of the piezoelectric scanner of the AFM, over a distance of 200 nm and at varying frequencies. As the static friction between the gold island and the graphite substrate is significantly smaller than the static friction between the tip apex and the island, this results in relative motion between the island and the substrate. The associated, time-dependent interfacial friction force is then directly measured by monitoring the torsional twisting of the AFM cantilever by way of laser beam deflection (*16*). Considering the miniscule values of friction force that arise at the superlubric gold-graphite interface that approach instrumental detection limits, it is of utmost importance to minimize the effect of noise on the acquired data. Motivated in this fashion, we employed a heterodyne detection approach based on a lock-in amplifier to collect the speed-dependent friction force data presented here (Fig. 2). This resulted in nearly an order of magnitude increase in our signal to noise ratios while recording friction forces.

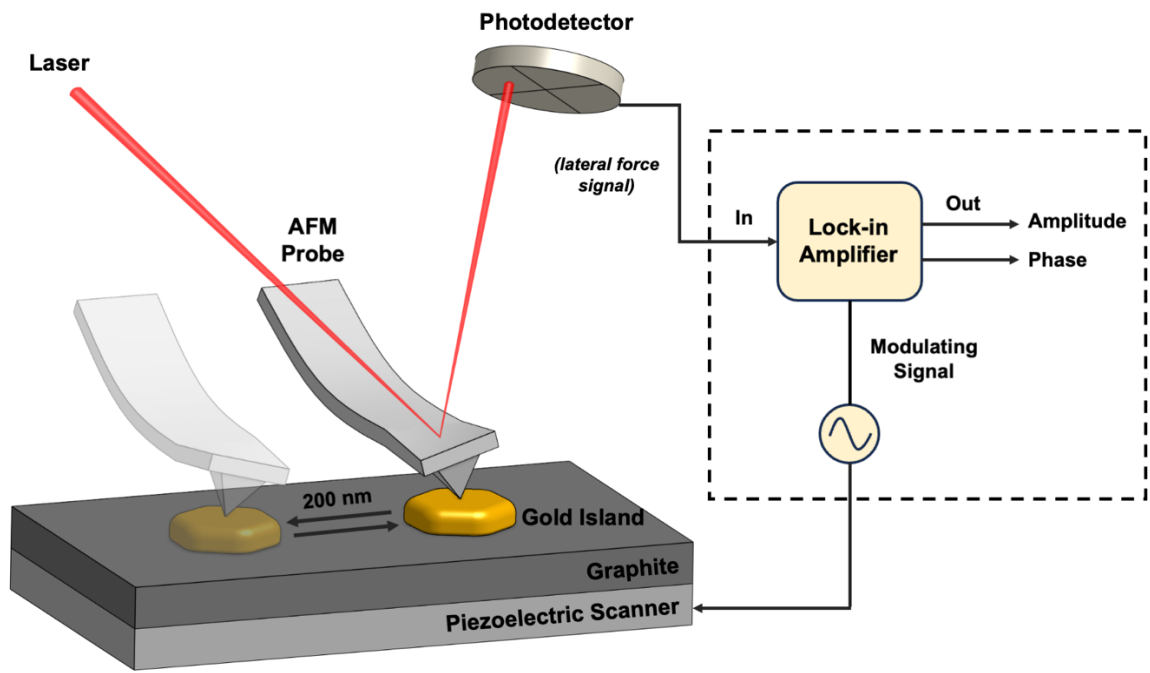

**Fig 2. Schematic of sliding experiments.** The AFM tip drags the gold island back and forth on the graphite substrate with a 200 nm stroke in a direction perpendicular to the long axis of the cantilever. A lock-in amplifier is used to extract the magnitude of the friction force from the twisting of the cantilever measured by the four-quadrant photodetector.

Representative results from our experiments are shown in Fig. 3. Fig. 3A shows the initial relocation step, where we move nanoislands of choice away from step edges, on which they are frequently located, to the "center" of an open graphite terrace prior to the sliding experiments. This relocation step is performed by first contacting the top surface of a nanoisland and then precisely manipulating it based on a reference line path. Once a targeted nanoisland is in a suitable location with the AFM tip in contact with its top surface, the sample is laterally modulated at varying frequencies, effectively sliding the top surface of the graphite substrate back-and-forth against the bottom surface of the nanoisland over a distance of 200 nm (Fig. 3B). These sliding experiments were repeated for a total of 10 different islands. In addition, two control experiments were conducted for each island. In the first control (indicated as Control I in Fig. 3C), we keep the AFM tip engaged on top of the nanoisland without any lateral modulation, while stepping through the same frequencies tested in the actual sliding experiments for data collection. The measurements from this control effectively describe the noise that couples to the lateral force signal while the AFM tip simply sits engaged on top of a nanoisland. In the second control (indicated as Control II in Fig. 3C), we repeat data collection through the same frequencies while the tip is free in air (i.e., disengaged from the nanoisland), a distance of 70 nm away from the sample surface. The measurements from this control describe the noise that couples to the lateral force signal while the tip is not interacting with the sample, such as the movement of the laser interference pattern during the sample modulation. Fig. 3C shows a plot of friction force versus sliding speed for a representative nanoisland, where the data from the actual island manipulation process (black) is plotted together with the data from the two control experiments (red and blue). Remarkably, friction increases by only a modest amount (from 52 pN to 68 pN) despite an increase in sliding speed that spans nearly three orders of magnitude. Additionally, even at the highest sliding speed of 115 μm/s where the total noise contributions are at their highest level, the actual lateral force signal is significantly higher, resulting in high signal-to-noise ratios. These results demonstrate the reliability of the acquired friction data, which allows us to discuss the speed dependence of structural superlubricity inferred from our results with confidence. Experiments performed on 9 additional islands (Fig. 4) provide similar results, whereby friction increases remain insignificant, with the largest relative increase exhibited by Island 3 (from 11 pN to 33 pN over the entire speed range).

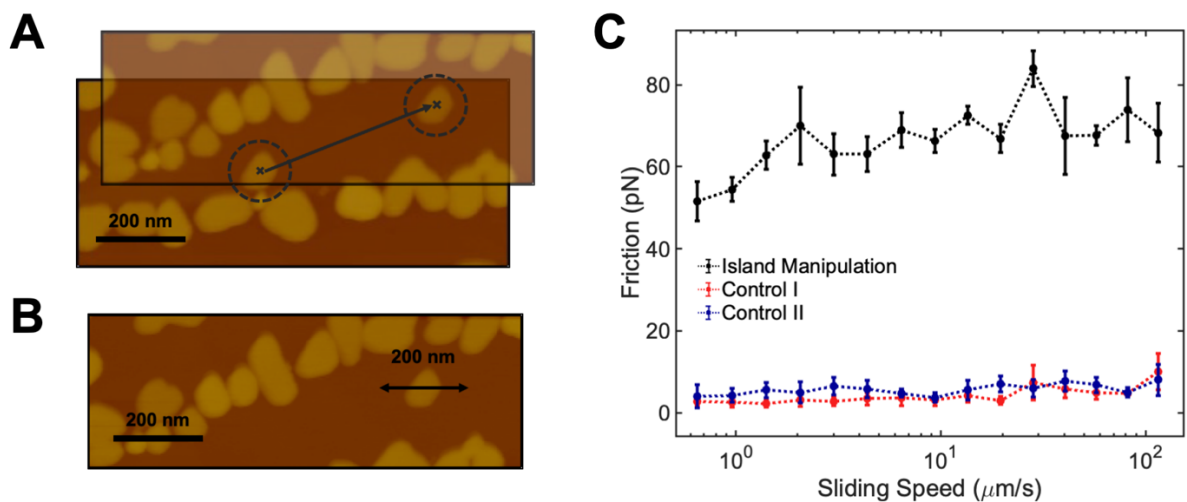

**Fig 3. Experimental procedure and representative results.** (**A**) Overlapping AFM topography images showing the relocation of a targeted nanoisland from a step edge to a graphite terrace. (**B**) AFM topography image of the targeted nanoisland from part (A), whereby the arrow indicates lateral modulation over a distance of 200 nm. (**C**) Representative plot of friction force versus sliding speed for a gold nanoisland, showing only a modest increase in friction force over three orders of magnitude of sliding speed (black). Noise contributions, as evaluated from two control experiments (red and blue), are significantly lower in magnitude than the friction force signal.

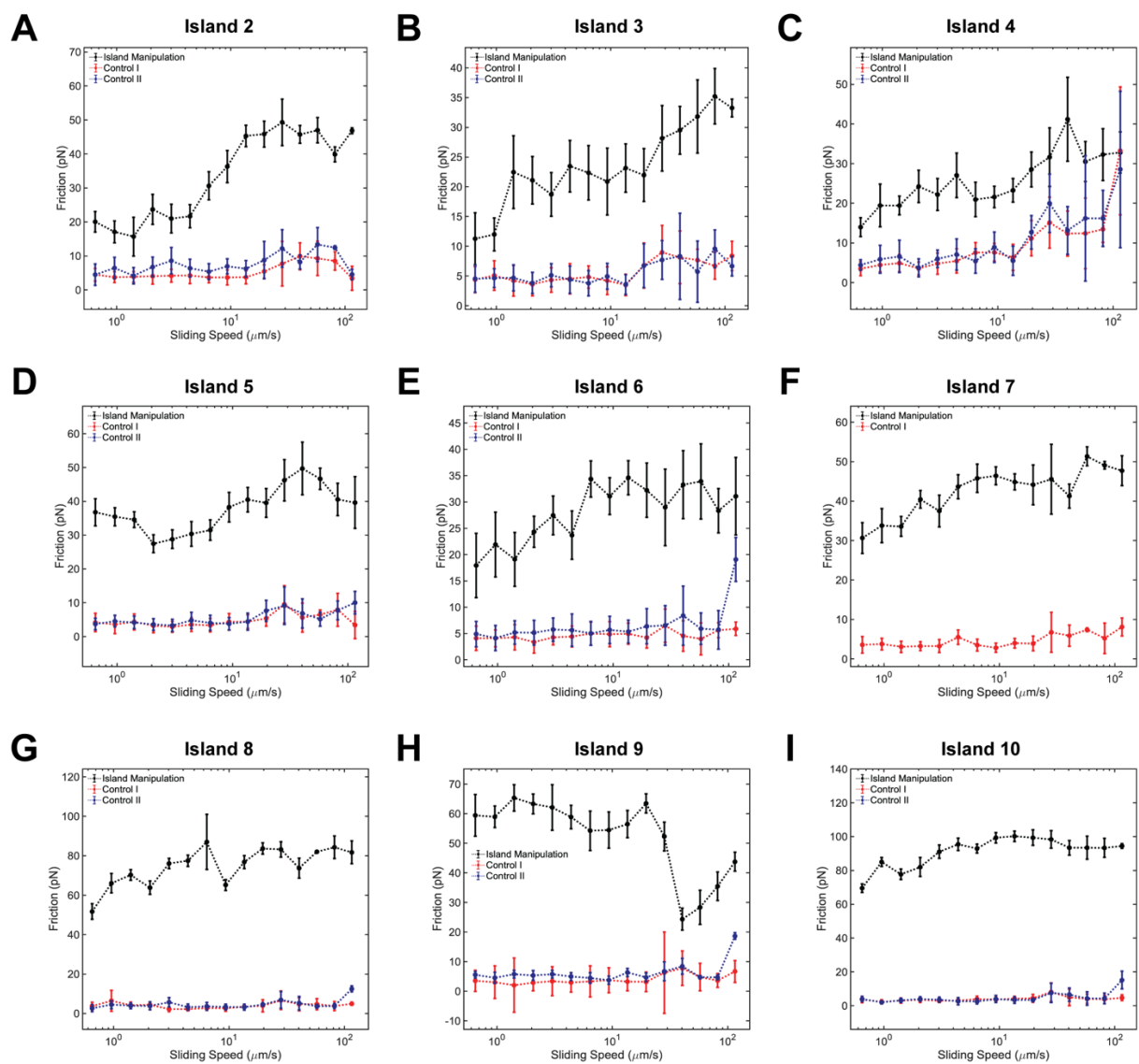

**Fig. 4.** Plots of friction force versus sliding speed for 9 additional nanoislands studied in our experiments.

Literature includes a limited number of studies aimed at theoretically predicting the effect of increasing sliding speeds on friction forces observed in structural superlubricity. In particular, it was claimed that friction at a molecularly clean, structurally superlubric contact would increase linearly (i.e., in a Stokesian fashion) with sliding speed (*11–13*, *17*), as opposed to the logarithmic dependence that is typically observed in conventional AFM-based measurements of speed-dependent friction based on the thermally-activated Prandtl-Tomlinson model (*18*). On the other hand, Gao *et al.* recently found evidence of velocity-insensitive (i.e., Coulombic, as opposed to Stokesian) friction for a gold slab sliding on graphite in a direction perpendicular to the contact line that is formed between the gold slider and graphite substrate (*19*). This observation was attributed to mechanical instabilities in the form of puckering of graphite under the slab near the propagating edge, creating moiré patterns that occasionally show abrupt, isolated disruptions. It is important to note that despite the mechanical instabilities leading to Coulomb-like frictions, the simulated shear stresses remain very low (<20 kPa), within the range expected for structurally superlubric contacts (*19*). Remarkably, the simulated stress values are in close alignment with the shear stress values we observe in the experiments presented here (<60 kPa) (*19*).

Our experimental results clearly demonstrate the absence of a significant increase in friction forces with sliding speed, aligning with the key characteristic of Coulombic friction. It should be indicated that the very mild yet statistically meaningful increase in friction force experienced by the majority of the nanoislands (with the exception of Islands 5, 6 and 9 where friction forces effectively remain constant within the error bars associated with our measurements) rule out the rather unlikely scenario of a complete absence of speed-dependent energy dissipation mechanisms. Rather, the presence of Coulombic as opposed to Stokesian friction implies that mechanical instabilities exist at the sliding interface (*19*, *20*). These can result from various factors including the graphite puckering issue indicated above as well as molecular contamination at the interface. In particular, theoretical work performed by Müser indicated that the presence of contaminant molecules at the sliding interface could lead to a switch from a linear (i.e., Stokesian) to a suppressed (i.e., Coulombic) speed dependence in friction force (*21*). Although we took careful measures to produce clean samples (via thorough cleaning of the AFM sample holder and the thermal evaporator prior to the experiments, as well as limiting thermal evaporation times and performing all measurements within 4 days after sample synthesis), and large-scale tapping-mode AFM images show no signs of contamination (as opposed to samples exposed to ambient conditions for longer durations, see Fig. S2), it is conceivable that some degree of contamination consisting of water and hydrocarbon molecules could still be present at the gold-graphite interface (*22*), potentially leading to the strongly suppressed speed dependence behavior we robustly observe in our experiments as an alternative and/or complementary mechanism to graphite puckering.

While we have no explicit control over various additional factors including the unavoidable presence of structural defects at the interface (*23*), our experiments demonstrate that the ultra-low friction state associated with structural superlubricity is preserved at high sliding speeds and under uncontrolled environmental conditions, bringing us one step closer to the realization of small-scale superlubric mechanical devices with minimal energy dissipation.

**Acknowledgements**

*Funding*

National Science Foundation grant 2131976 (MZB)

National Science Foundation Graduate Research Fellowship Program (WHO)

Office of Science, Basic Energy Sciences Program of the U.S. Department of Energy grant DE-AC02-05CH11231 (PDA)

**Supplementary Materials**

*Materials and Methods*

The gold-graphite material system utilized during the measurements was prepared by thermal evaporation of small amounts (~2 nm) of 99.999% purity gold on freshly cleaved highly oriented pyrolytic graphite (HOPG) substrates under high vacuum (~$10^{-8}$ Torr). The deposition step was conducted at room temperature (~20 °C) for time periods in the range of 30 – 90 seconds. Subsequently, the sample was annealed at elevated temperatures (in the range of 250 – 300 °C) to promote the aggregation of gold clusters into nanoislands. This process resulted in a sample system with crystalline gold islands of various size (1000 – 15000 nm$^2$) on the HOPG surface, with the majority of nanoislands decorating step edges and some nanoislands located on flat terraces (see Fig. S1).

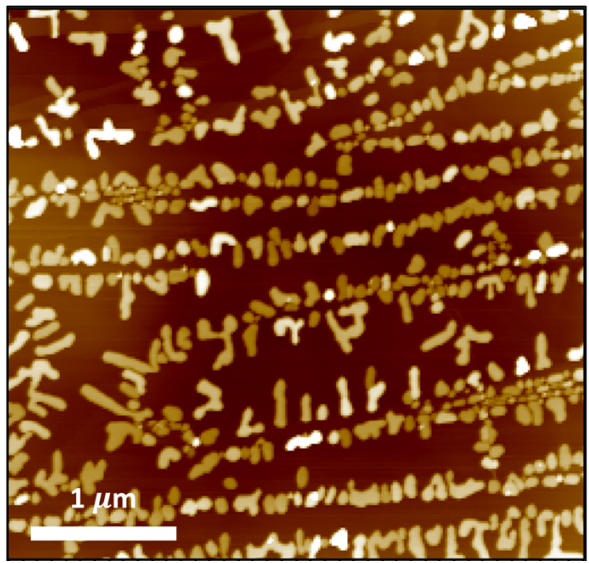

**Fig. S1.** Large-scale topography image of the gold-graphite sample.

Before conducting the sliding experiments, we took careful measures to produce and keep samples as clean as possible. Firstly, we identified two major sources of contamination: thermal evaporator and AFM. To minimize contamination from the evaporator, we cleaned and wiped down the surfaces of the bell cage, used a new boat, modified the sample recipe to minimize time in evaporator, and annealed samples at atmospheric pressure to limit adsorption of contaminants. To minimize contamination from AFM, we cleaned the AFM holder and stage (by sonicating the parts in acetone and ethanol for a few minutes) and cleaned the AFM tips with ozone. Lastly, to maintain the freshness of the samples, we conducted all our sliding experiments within 4 days after sample synthesis. In AFM topography and phase images, we observed no signs of contamination in our fresh samples (Fig. S2 B and D), in contrast with our contaminated, old samples (Fig. S2 A and C).

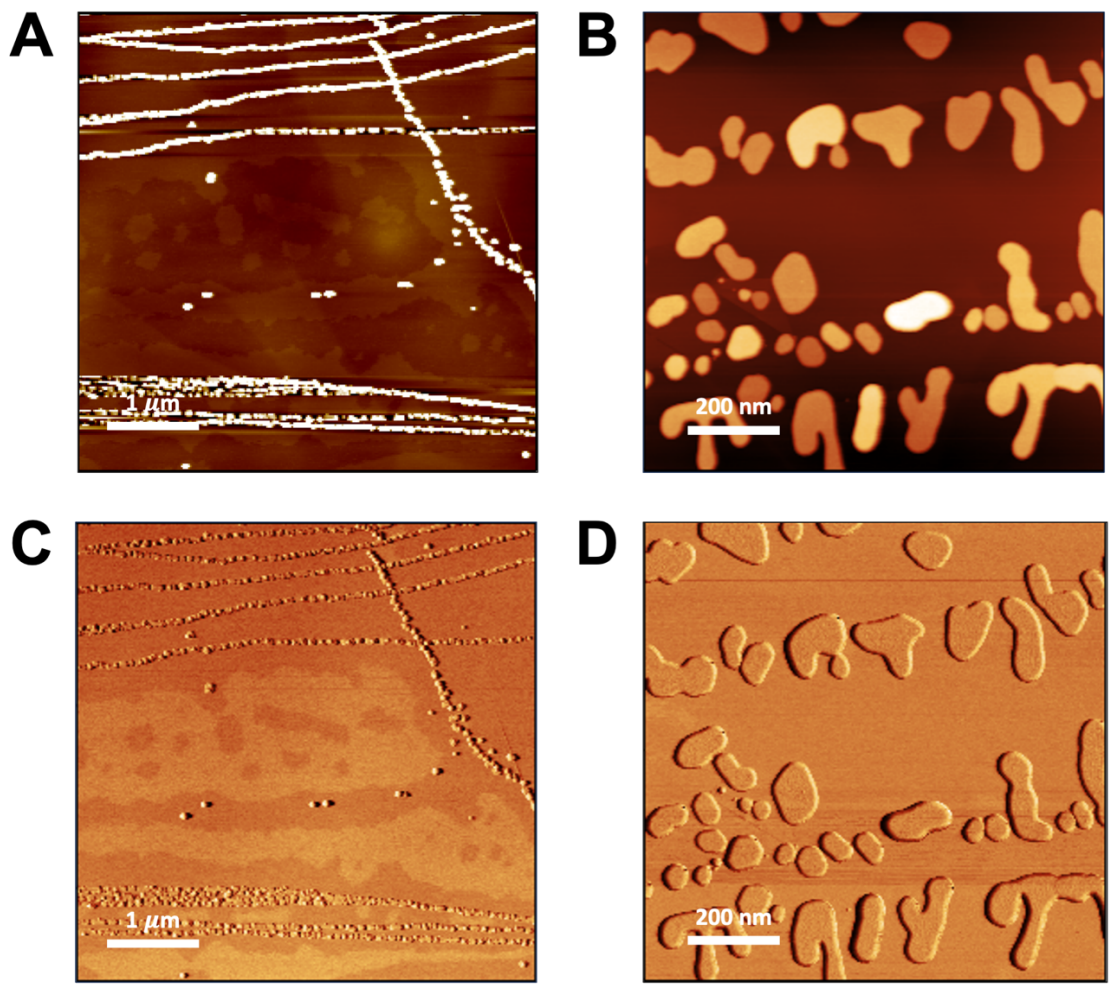

**Fig. S2.** Comparison of contaminated and clean samples. (A) Topography image of a clearly contaminated sample, recorded several weeks of exposure to ambient conditions after synthesis. (B) Topography image of the apparently clean sample used for all sliding experiments presented here. (C) Phase image recorded concurrently with the topography image in A, showing extensive "puddles" of contamination. (D) Phase image recorded concurrently with the topography image in B, showing the absence of contamination "puddles".

Lateral manipulation experiments were performed on the gold-graphite material system fabricated as above by way of the "tip-on-top" method, utilized in conjunction with a commercial AFM instrument (Asylum Research, Cypher VRS) and cantilevers (Budget Sensors, Tap150Al-G) and an external lock-in amplifier (Zurich Instruments, HF2LI). The normal and lateral calibration of the cantilevers have been performed via the Sader method (*24*) and the wedge calibration method (*25*), respectively. An initial frequency sweep of the *y*-sensor piezo revealed a resonance peak at 280 Hz (Fig. S3), which we avoided by limiting our lateral modulation of the sample system only up to 240 Hz.

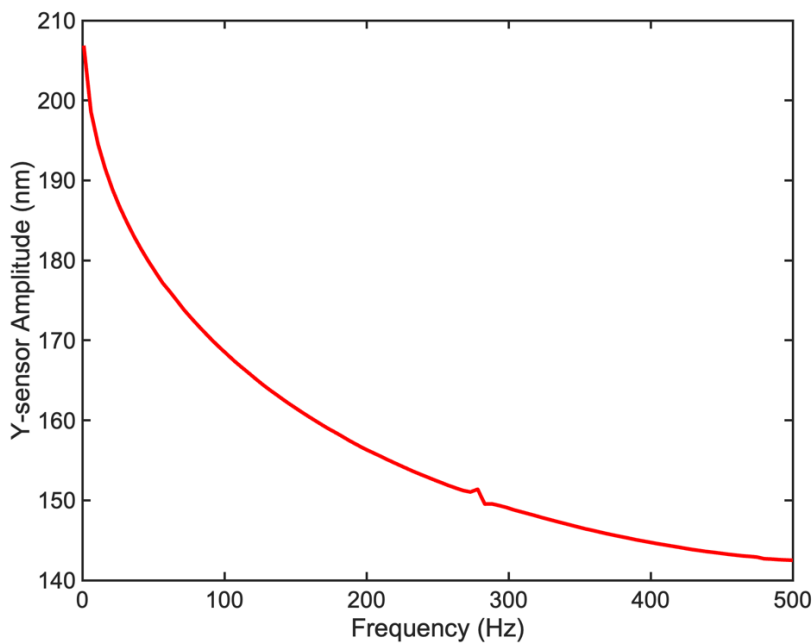

**Fig. S3.** Initial frequency sweep of the *y*-sensor piezo reveals a resonance peak at 280 Hz.

The experimental procedure was performed by the following steps: (i) First, a suitable island was located via tapping-mode imaging, (ii) the targeted island was relocated from its initial position (typically near step edges and defects) to an open graphite terrace with the AFM "lithography" tool, (iii) a force curve was taken on top of the island surface right before the sliding experiments, (iv), the sample was modulated 200 nm back and forth at varying frequencies (from 1 – 240 Hz) using a lock-in amplifier, (v) a force curve was taken on top of island right after the sliding experiments. Any drift in the $z$-position during the sliding experiments (e.g., caused by the tip releasing the island during the modulation) could be detected by a significant shift of the baseline of the force curve would (see Fig. S4) and the friction force signal would drop to noise levels (i.e., ~5 pN).

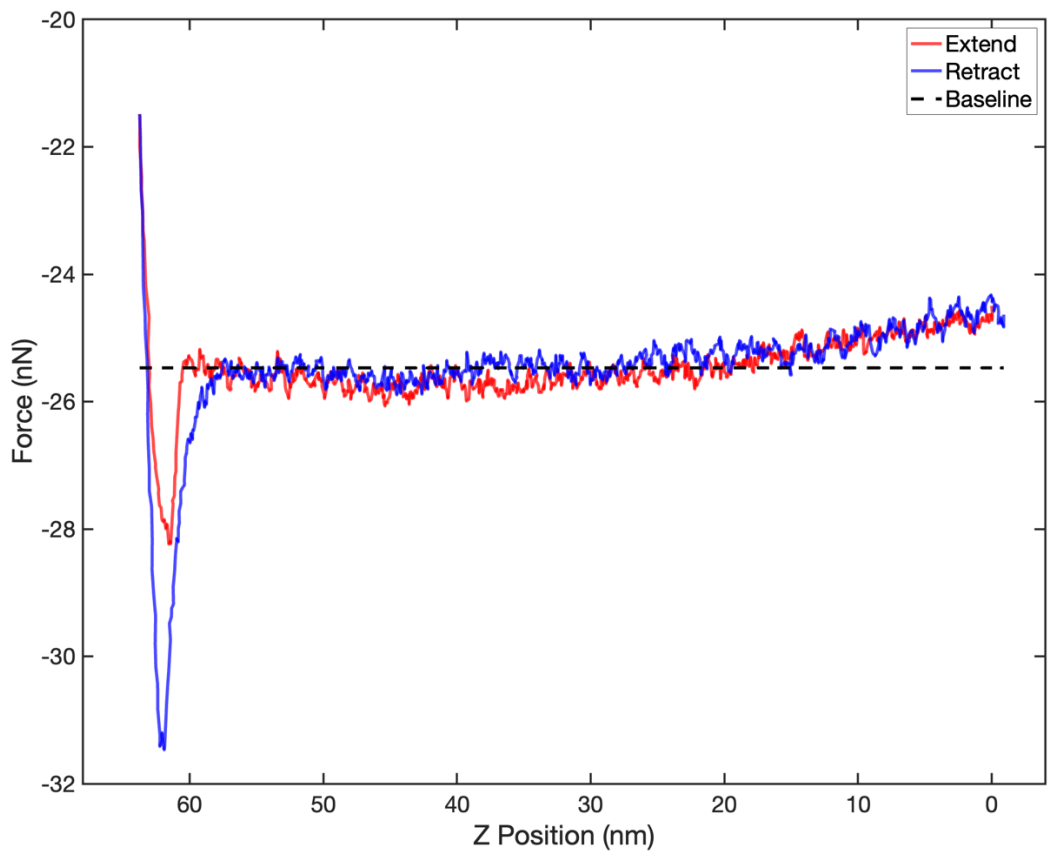

**Fig. S4**. Representative force-distance curve taken on top of an island surface.

*Supplementary Experimental Results*

Fig. S5 is a representative phase vs. speed plot from the experiments. The small error bars (standard deviation) of the phase signal show that the system was properly "locked in" by the lock-in amplifier during the experiments, increasing our confidence in the reliability of our measurements. The significantly large error bars of the controls confirm that our control measurements were pure noise. Additionally, we explored the effect of load on the speed-dependent sliding experiments. As shown in Fig. S6, additional normal load does not appear to affect the friction values or the friction trends with speed.

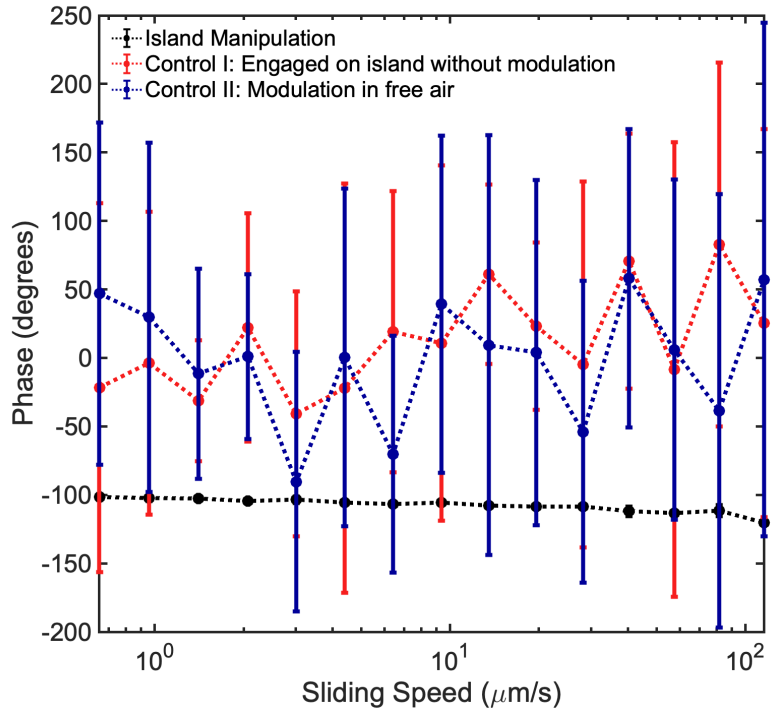

**Fig. S5.** Representative phase vs speed plot from sliding experiments.

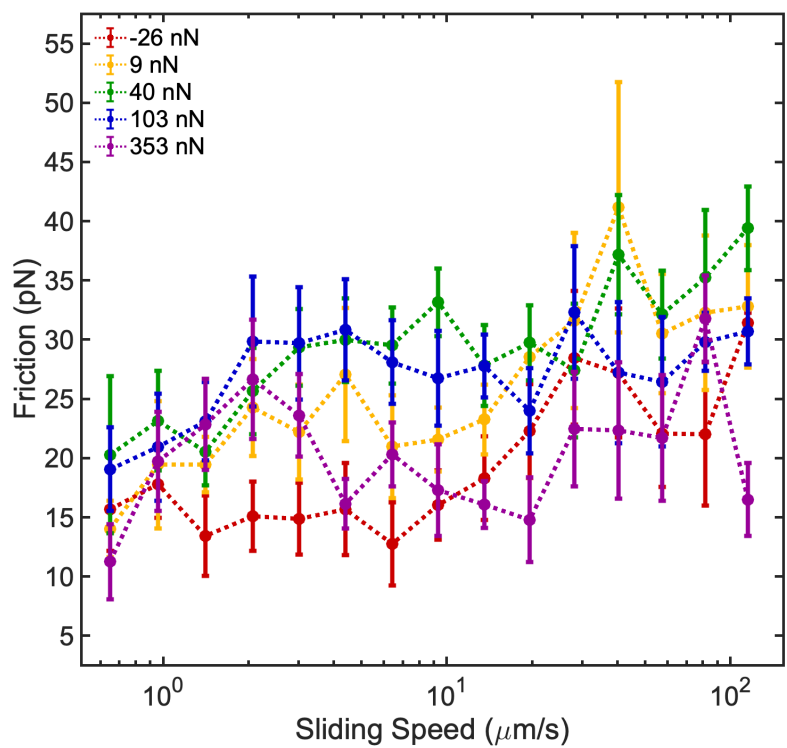

**Fig. S6.** Plot of friction force versus sliding speed for Island 3, where load dependence was tested. There is no clearly discernable effect of load on the friction forces or the speed dependence trends, up to the applied maximum of 353 nN.

23